\documentclass[
twocolumn,
twoside,
letterpaper
]{IEEEtran}

\ifCLASSINFOpdf
  \usepackage[pdftex]{graphicx}
  \DeclareGraphicsExtensions{.pdf,.jpeg,.png}
\else
\fi
\usepackage{amsmath, bm, mathabx}

\usepackage[colorlinks,citecolor=red,urlcolor=blue,bookmarks=false,hypertexnames=true]{hyperref}

\usepackage{color}
\definecolor{brickred}{rgb}{0.8, 0.25, 0.33}

\begin{document}
%
\title{\LARGE\bf Characterization of NbTiN  films with thicknesses below 20\,nm for low power kinetic inductance amplifiers}

\author{A.~Giachero, 
        M.R.~Vissers, 
        J.D.~Wheeler,
        M.~Malnou, 
        J.E.~Austermann, \\
        J.~Hubmayr, 
        A.~Nucciotti, 
        J.N.~Ullom, 
        J.~Gao
\	   
\thanks{A.~Giachero 
        is with Dipartimento di Fisica, Universit\`{a} di Milano-Bicocca, , also with
        INFN - Sezione di Milano Bicocca, Milano, I-20126, Italy, also with
        Department of Physics, University of Colorado, Boulder, Colorado 80309, USA, and also with 
        National Institute of Standards and Technology, Boulder, Colorado 80305, USA
        (e-mail: andrea.giachero@nist.gov).
        }
\thanks{M.~Malnou,
        J.N.~Ullom, 
        J.~Gao
        are with Department of Physics, University of Colorado, Boulder, Colorado 80309, USA,
        also with
        National Institute of Standards and Technology, Boulder, Colorado 80305, USA
       }
\thanks{M.R.~Vissers,
        J.D.~Wheeler,
        J.E.~Austerman,
        J.~Hubmayr
        are with National Institute of Standards and Technology, Boulder, Colorado 80305, USA 
        }
\thanks{A.~Nucciotti  
        is with Dipartimento di Fisica, Universit\`{a} di Milano-Bicocca, 
        and also with
        INFN - Sezione di Milano Bicocca, Milano, I-20126, Italy}
}


\maketitle

\begin{abstract}

A quantum-limited amplification chain is a fundamental advantage for any application that may benefit from the detection of very faint signals. Reading out arrays of superconducting detectors (TESs or MKIDs), resonant cavities, or qubits, calls for large bandwidth amplifiers in addition to having the lowest possible noise. At millikelvin temperatures, Kinetic Inductance Traveling-Wave Parametric Amplifiers (KI-TWPAs) working in 3-wave-mixing (3WM) and fabricated from a 20 nm thick NbTiN film have shown promising noise performances, as they can operate close to the quantum limit~\cite{Malnou2021}. However, they still require fairly high pump power. Devices that would require lower pump power would be easier to implement in readout chains, could reach the quantum limit and they would be compatible with qubit readout. A possible solution for obtaining this optimal configuration is to use a thinner superconducting film. In this work we explore the properties of  NbTiN films with a thickness less than 20 nm and we report the obtained experimental characterizations in terms of critical temperature, normal resistivity, and kinetic inductance. A new design for a 3WM KI-TWPA amplifier, based on these developed superconducting films, is introduced and discussed.
\end{abstract}


%
\IEEEpeerreviewmaketitle

\section{Introduction}\label{sec:intro}
Ultra-sensitive detection schemes play a central role in many advanced applications, mostly for fundamental physics experiments, such as: detection of axionic dark matter~\cite{Braggio2022}, dark photons~\cite{Ramanathan2022}, astrophysical measurements~\cite{Zobrist2019}, fundamental physics experiments~\cite{Giachero2022} but also are key factors for the deployment of quantum technologies, and in particular for the qubit readout~\cite{Ranzani2018, Heinsoo2018}, quantum key distribution~\cite{Fesquet2022}, and microwave quantum illumination~\cite{Fasolo20221}. In many of these applications, the necessity of reading a large array of detectors, cavities, qubits, calls for large bandwidth amplifiers with the lowest possible noise. One of the leading proposals for achieving these requirements is through the use of a broadband traveling wave parametric amplifier (TWPA) such as Josephson JTWPA~\cite{Macklin2015} or kinetic inductance KI-TWPA~\cite{HoEom2012}. These newly developed devices can reach the noise level near the quantum limit with high gain over a larger bandwidth (a few GHz). TWPA amplifier generally consists of a transmission 

KI-TWPAs achieve amplification through wave-mixing processes induced by the film's intrinsic nonlinear superconducting kinetic inductance. When a strong pump current propagates with a weak signal current along a line, energy is transferred from the pump to the signal, achieving a parametric amplification. 
For these films, the kinetic inductance per unit length $L_k$ of the transmission line can be expanded as a series of even powers of line current $I$ as follows:

 \begin{equation}\label{eq:kin}
    L_k(I)=L_0\left[1+\cfrac{I^2}{I_*^2} + \mathcal{O}(I^4)\right]
\end{equation}
where $L_0$ is the kinetic inductance per unit length in the absence of inductive non-linearity ($I=0$) and $I_*$ controls the scale of the non-linearity. From the Mattis-Bardeen~\cite{Mattis1958} $L_0=\hbar\, R_{n}/(\pi\Delta)$ with $\Delta=1.76\,k_{\scalebox{0.6}{\mbox{B}}}\,T_c$, where $R_{n}$  is the normal state resistance, $T_c$ is the critical temperature and $k_{\scalebox{0.6}{\mbox{B}}}$ the Boltzmann constant. $I_*$ is of order of the critical current $I_c$ of the film ($I_*=3\sqrt{3}/2\,I_c$~\cite{Zmuidzinas2012}) and depends on the superconducting material and geometry.

By the introduction of a DC bias applied to the input of the amplifier the inductance equation becomes:
\begin{equation}\label{eq:kinbias}
    L_k(I)=L_d\left(1+ \varepsilon I+\xi I^2 \right)\,\,\mbox{with}\,\,L_d=L_0\left(1+\cfrac{I_{dc}^2}{I_*^2}\right)
\end{equation}
where $I_{dc}$ is the applied current, $L_0$ is the NbTiN kinetic inductance,  $L_d$ is the amplifier inductance per unit under DC bias at zero RF current ($I=0$), $\varepsilon=2I_{dc}/(I_*^2+I_{dc}^2)$, and $\xi=1/(I_*^2+I_{dc}^2)$~\cite{Malnou2022}. 
The term $\xi I^2$ permits the input pump and signal to 4-wave-mix (4WM), producing an idler at $f_i=2f_p-f_s$, while the term $\varepsilon I$ depends on the DC bias and permits the 3-wave-mix (3WM), producing an idler at $f_i=f_p-f_s$. In 4WM operating mode $L_k$ depends exclusively on the RF pump power ($P\propto I^2$) while in 3WM $L_k$ depends on both the RF pump power and the DC bias ($I_{dc}^2$). 
An important consequence of this is that in 3WM less RF pump power is required to generate RF mixing~\cite{Visser2016}.

\begin{figure}[!t]
\centering \includegraphics[width=0.3\textwidth,clip]{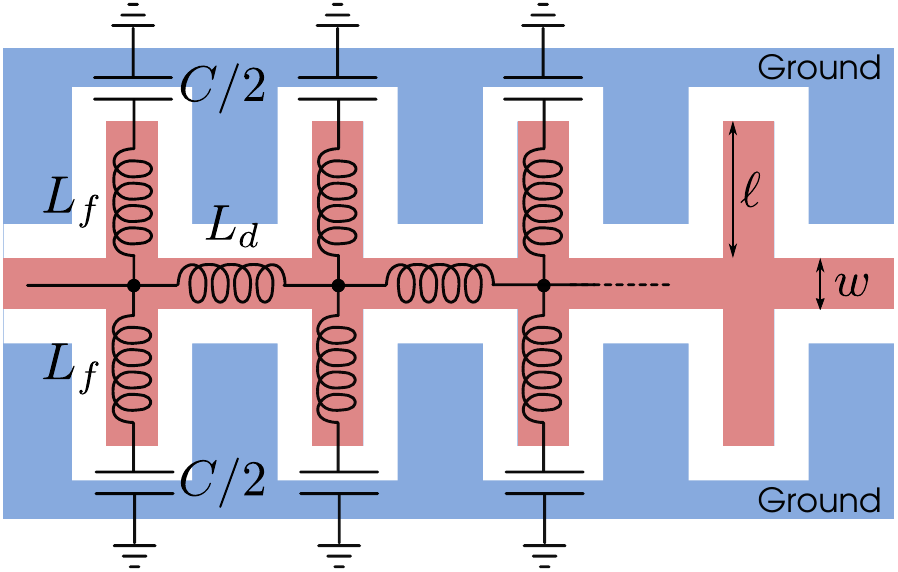}
\caption{Schematic of the KIT amplifier artificial transmission line with three cells in series (not to scale) are arranged in a CPW topology.}
\label{fig:kitdes}
\end{figure}

KI-TWPA amplifier working in 3WM and fabricated at NIST from a 20\,nm thick NbTiN film has shown promising performance~\cite{Malnou2021,Malnou2022}. The developed design is composed of a long transmission line  consisting of several elementary cells. Each cell consists of a coplanar waveguide (CPW), with inductance $L_d$, and two interdigitated capacitor (IDC) stubs, one for both sides of the center line, that form the capacitance to ground $C$ (figure \ref{fig:kitdes}). The stub length $\ell$ is tuned in order to obtain a $C$ value such that $Z_0 = \sqrt{L_d/C} = 50\,\Omega$~\cite{Chaudhuri2017}. In order to generate the exponential parametric amplification, a dispersion engineering has been accomplished using periodic loadings created by decreasing the length of the stubs such that $Z_0 = \sqrt{L_d/C} = 80\,\Omega$, as explained in \cite{Chaudhuri2017} and \cite{Malnou2021}  

Characterization results showed an average gain of 16.5\,dB from 3.5 and 5.5 GHz, and a typical 1\,dB compression power of -63\,dBm at the amplifier input within that bandwidth~\cite{Malnou2021}. The measured system-added noise resulted in $N_\Sigma=3.1 \pm 0.6$ quanta (equivalent to $T_N=0.66 \pm 0.15$\,K), suggesting that the KI-TWPA amplifier alone operates near the quantum limit. This performance has been obtained with an intrinsic scaling current of $I_*=7$\,mA, a bias current of $I_{dc}=1.5$\,mA, and a pump signal with a power of -28\,dBm at the input of the device. Despite these very promising results, the developed device still requires a fairly high pump power to reach the maximum gain. The pump must be isolated from the device under test by components that unavoidably insert loss, thereby degrading the noise performance of the chain. An amplifier functioning with a lower pump power may necessitate fewer of these nefarious isolating components and may improve the read out performance. 

One solution to overcome this problem is to use thinner superconducting films. In fact, decreasing the thickness $t$ of the film increases both the kinetic inductance $L_0$ along with the inductive non-linearity. In addition, since $I_*$ is proportional to the cross-sectional area of the film $A=t\cdot w$, where $w$ is the inductor width, this means that by lowering thickness $t$ the scaling current $I_*$ decreases. Considering equation \ref{eq:kinbias} it is possible to see that with a higher $L_0$ and a lower $I_*$ it is possible to obtain the same inductive non-linearity with a lower pump power $P\propto I^2$ and lower bias current $I_{dc}$. 
    
Before designing a KI-TWPA amplifier based on thinner NbTiN films the complete knowledge of the material and of its intrinsic parameters is a fundamental requirement. To investigate the thickness dependence of the properties of the film, we fabricated NbTiN films with different thicknesses $t=5, 10, 20$\,nm. The $20$\,nm thickness, commonly used for the development of KI-TWPA amplifiers at NIST, was used as a comparison since its intrinsic properties are widely known. Following the developed fabrication procedure, the superconducting films are deposited on a high-resistivity, intrinsically doped, float-zone Silicon wafer. Immediately before loading into the deposition chamber, the wafer is etched with hydrofluoric acid to remove the native oxide~\cite{Wisbey2010}. The NbTiN is reactively co-sputtered from Nb and Ti targets in an Ar-N$_2$ atmosphere at 500\,$^{\circ}$C.  Similarly to the thicker parametric amplifier films, the deposition rates of the reactively sputtered NbN and TiN are tuned to maximize the $T_c$ for the co-deposited alloy~\cite{Bockstiegel2014}.  After deposition, the films are lithographically patterned using an i-line stepper and are dry-etched in an ICP-RIE with a CHF$_3$ based plasma. This etch leaves an etched Si surface that is compatible with high-Q microwave devices but with minimal trenching into the Si that could result in a spatially varying impedance across the wafer. 

For each thickness, we extrapolated the fundamental material properties by performing DC measurements on superconducting test structures, and RF measurements on superconducting resonators.

\begin{figure}[!t]
\centering \includegraphics[width=0.45\textwidth,clip]{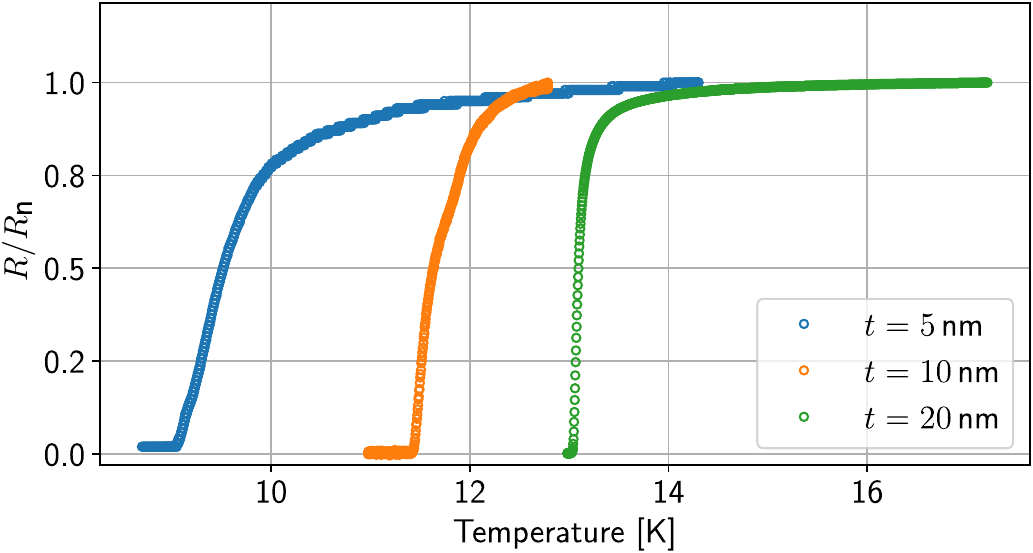}
\caption{Normalized Sheet resistance vs temperature indicating a transition from a metallic normal state to the superconducting state for the different film thicknesses. Each transition curve has been measured in a different cool-down in a range of temperatures around the film's critical temperature.}
\label{fig:Tctrans}
\end{figure}

\begin{table}[!t]
\caption{Kinetic inductance and sheet resistance result from 4-terminal sensing measurements \label{tab:kinDC}}
\centering
\begin{tabular}{ccccc}
\hline\hline
$\bm t $  & $\bm T_c$ & $\bm R_{\scalebox{0.6}{\mbox{n}}}$ & $\bm \rho$   & $\bm L_k$ \\ 
$ $[nm]    & [K]       & [$\Omega/$sq]                    & [$\Omega\cdot$nm] & [pH/sq] \\  \hline
5         & 9.5  & 817 & 4089 & 118.6 \\ 
10        & 11.6 & 277 & 2772 & 32.9  \\
20        & 13.1 & 100 & 2002 & 10.5  \\ 
\hline\hline
\end{tabular}
\end{table}

\section{Kinetic Inductance extracted from resistances measurements}\label{sec:dc}
From the Mattis-Bardeen relationship, as discussed in the preceding section, it is clear that the kinetic inductance increases as the sheet resistance increases (or the critical temperature decreases). In order to measure the critical temperature and the sheet resistance, the electrical resistances of the film's test structure were measured with the 4-terminal sensing techniques. The transition curves as a function of the temperature are reported in figure \ref{fig:Tctrans} and the obtained results for the extrapolated characteristic parameters are reported in table \ref{tab:kinDC}. From these measurements, it is possible to see that as the film becomes thinner, $T_c$ decreases and $R_{n}$ increases. This means that also $L_k$ increases, as expected. The kinetic inductance measured for the 20\,nm  film is compatible with previous results ($\sim 10$\,pH/sq~\cite{Malnou2021}) while for 10\,nm and 5\,nm the obtained values are around 3 times and one order of magnitude larger, respectively.

\section{Kinetic Inductance extracted from resonator characterizations}\label{sec:rf}
\begin{figure}[!t]
\centering \includegraphics[width=0.45\textwidth,clip]{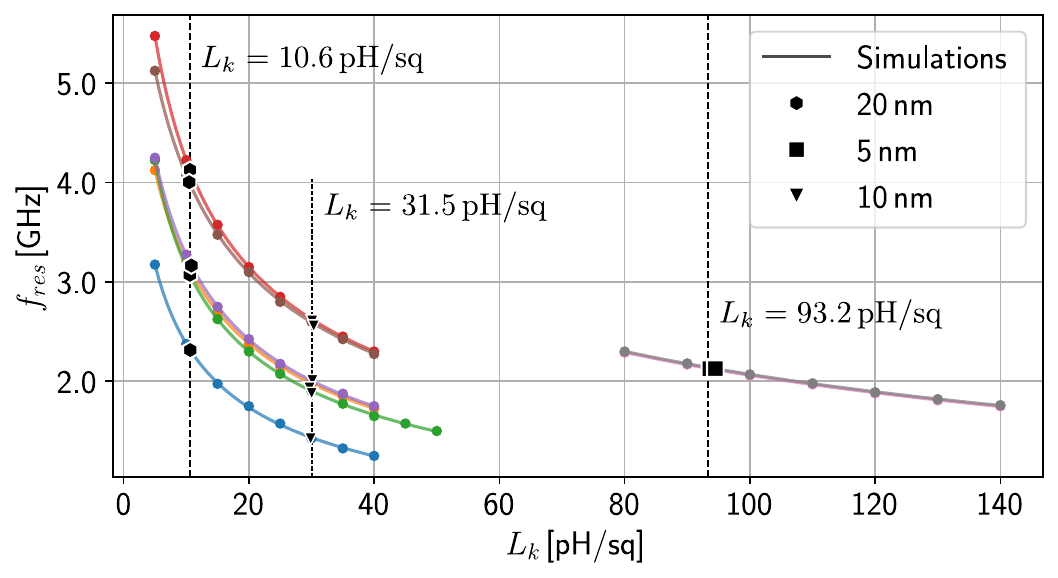}
\caption{Calibration curves for the simulated resonant frequencies (solid colored lines with filled circle points) as a function of the kinetic inductance compared with the measured values for the different resonators and film thicknesses (hexagon, square, and triangle filled points). The vertical dashed black line shows the  average kinetic inductance values for every thickness.}
\label{fig:kincal}
\end{figure}

To explore the thickness dependence, we have also fabricated lumped element LC resonators from NbTiN films. For each resonator, the LC resonance is defined by an interdigitated capacitor with $10\,\mu$m wide fingers and gaps and by a strip inductor 1 or  $2\,\mu$m wide, depending on the design. Starting from the $L_k$ values from the DC characterization, we designed and tested two different versions of resonator arrays, the former optimized for 20 and 10\,nm while the latter optimized for 5\,nm. The resonator chip arrays are designed with individual resonance frequencies distributed in the 1-4\,GHz range. For each resonator, we estimated the resonant frequency position as a function of the film kinetic inductance by exploiting Sonnet EM simulations~\cite{Wisbey2014}. This allows us to obtain a set of $f_{\scalebox{0.6}{\mbox{res}}} = f(L_k)$ calibration curves~\cite{Gao2014} (figure \ref{fig:kincal}, solid lines). Then $L_k$ values were extrapolated by comparing the measured resonance frequencies with the calibration curves (figure \ref{fig:kincal}, black markers). The obtained results are reported in table \ref{tab:kinRF} and in figure \ref{fig:kincal}. The resonance frequencies were measured using a vector network analyzer (VNA) at a bath temperature of 50\,mK. Each acquired resonance was fitted by using a python framework developed for microwave kinetic inductance detectors (MKID) for mm/submm astronomy~\cite{Wheeler2022}. 

\begin{table}[!t]
\caption{Kinetic inductance and internal qualities factor extrapolated for the different film thicknesses \label{tab:kinRF} at 50\,mk}
\centering
\begin{tabular}{ccc}
\hline\hline
$\bm t$  & $\bm L_k$ & $\bm Q_i$ \\  
$ $[nm]     & [pH/sq]   & [$10^3$]              \\  
\hline
5       & 93.3  & 30  \\
10      & 31.5  & 50   \\
20      & 10.6  & 70  \\
 \hline\hline
\end{tabular}
\end{table}

The difference between the kinetic inductance values obtained with the RF characterizations and the values calculated from the local Mattis–Bardeen’s relationship (section \ref{sec:dc}) is 1\% for 20\,nm the 4\% for 10\,nm, and the 21\% for 5\,nm. In the case of 20\,nm the test structures and resonator came from the same wafer and the difference is negligible. This is not true for 10\,nm and 5\,nm and the larger difference can be justified by the different fabrication runs. From past fabrications, we found a spread on the same wafer of 1-2\% in the film parameters (film thickness, critical temperature, and normal resistivity). In separate wafers but using the same process steps and device designs, the deviation was found to be less than 10\%. Further improvements on the fabrication procedure are in progress in order to minimize this effect. The obtained values for the kinetic inductance $L_k$ have the same order of magnitude and this is sufficient for designing a parametric amplifier, since the line finger length is designed to have less than $Z_0=50\,\Omega$", target value that is then achieved by tuning the current bias. 

For all the measured resonators, the internal quality factors $Q_{i}$ resulted larger than $3\cdot 10^{4}$ and this means an internal loss $1/Q_i$ below $3.3\cdot 10^{-5}$. As reported in~\cite{Malnou2022} Appendix B, considering a KI-TWPA amplifier composed by $\sim 6\cdot 10^6$ cells, any internal quality factor larger than $5\cdot 10^{3}$ ($1/Q_i<2\cdot 10^{-4}$) produces an insertion loss below $1$\,dB. This means that for KI-TWPA amplifiers based on the films characterized here, the device contribution to the total insertion loss would be negligible. 

\begin{figure}[!t]
\centering \includegraphics[width=0.45\textwidth,clip]{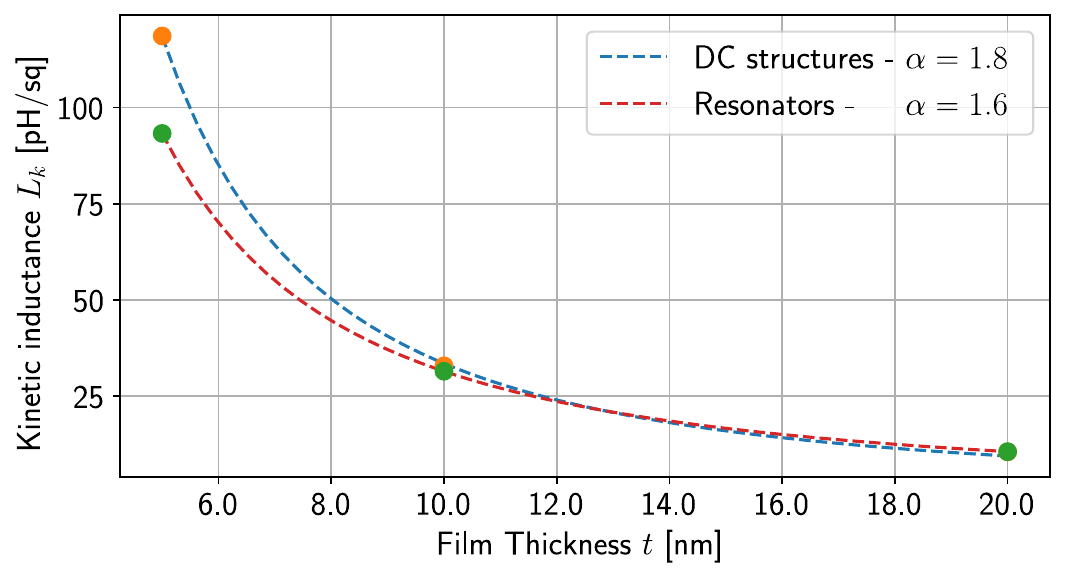}
\caption{Kinetic inductance as a function of the film thickness, for both the DC and RF characterizations. The dashed lines represent the best fits performed with a $\propto 1/t^\alpha$ model.}
\label{fig:kintick}
\end{figure}

In figure \ref{fig:kintick} the obtained $L_k$ values as a function of the film thicknesses $t$ are shown for both the DC and RF characterizations. The obtained trends allow us to roughly calibrate the fabrication process for obtaining the wanted $L_k$ by tuning the $t$.

\section{Amplifier design for thinner thickness}
The previous KI-TWPA amplifiers developed at NIST with films 20\,nm NbTiN thick, presented a kinetic inductance around $L_k=10$\,pH/sq~\cite{Bockstiegel2014, Visser2016, Malnou2021}. Considering this value of $L_k$, and considering the transmission line design presented in Section \ref{sec:intro} (figure \ref{fig:kitdes}), to have a capacitance $C$ to ground such that $Z_0=50\,\Omega$ the stub length had to be $102\,\mu$m long on both sides. If the kinetic inductance rises the $C$ value needed to match the $Z_0=50\,\Omega$ condition is necessarily obtained with larger $\ell$. By keeping the central line $w=1\,\mu$m and considering $L_k$ values of around $30$\,pH/sq ($t=10\,$nm) and $100$\,pH/sq ($t=5\,$nm) here obtained, this means, to first order, a stub length of 3 times and one order of magnitude longer, respectively, with respect to the 20\,nm version. Such stub lengths would not be compatible with the compact design introduced in~\cite{Malnou2021} in which $\sim 6\cdot 10^6$ cells fit in a $2\times 2$\,cm$^2$ chip by exploiting an optimized double spiral topology. In particular, longer stub lengths would mean less space available on the chip to pattern the device line, a larger radius for the curved part of the line, a larger arc in the spiral corners, and less repetition of the spiral patterns. All of these restrictions are translated into a shorter KI-TWPA transmission line that would produce a lower gain.          
   
\begin{figure}[!t]
\centering \includegraphics[width=0.46\textwidth,clip]{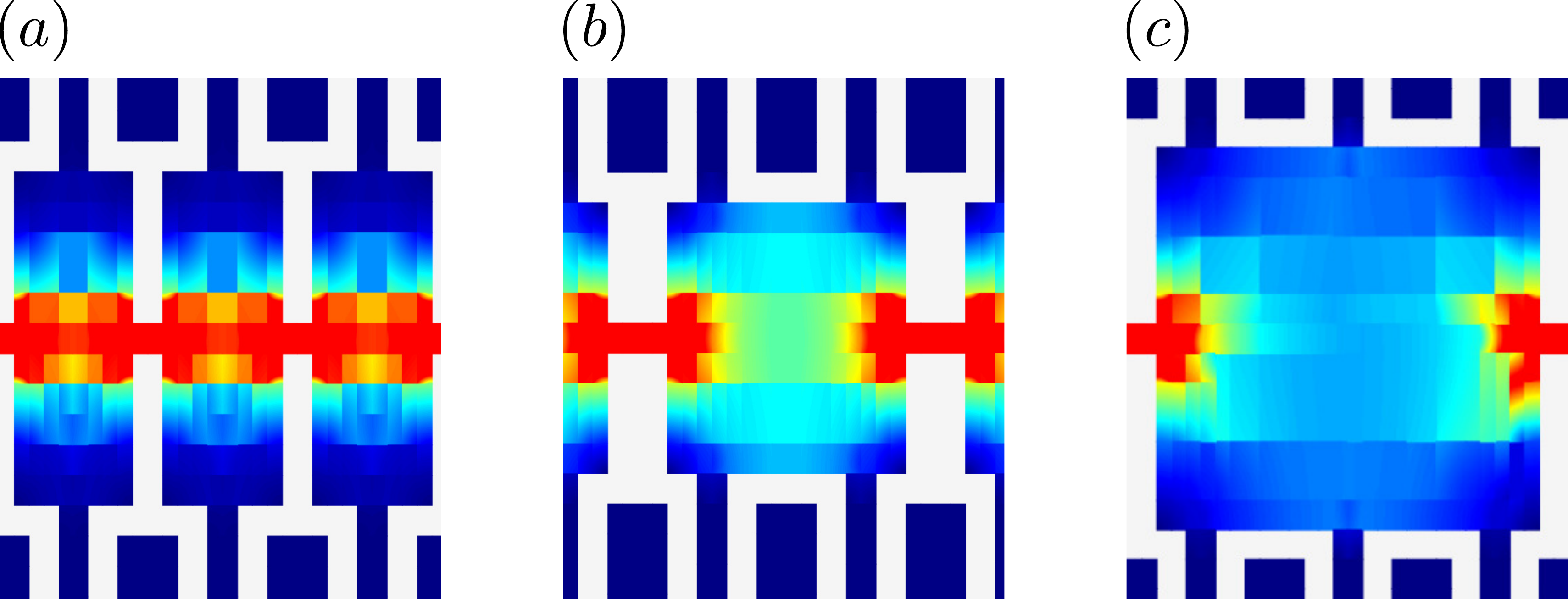}
\caption{The new KI-TWPA design with the central line is implemented as a series of wider and narrower parts. In the three different configurations, the wider part of the line covers $(a)$ one, $(b)$ two, and $(c)$ three stubs. The heat map shown in each configuration is related to the RF current density in the line during the simulations.}
\label{fig:wider}
\end{figure}
         
One way to keep short the stub lengths would be to enlarge the central line width $w$. This would reduce the number of squares in the central line, keeping low the inductance per cell, but, at the same time, since $I_*\,\propto\,w\cdot t$, This would result in an increase in the scaling current $I_*$, requiring a larger pump power to operate the KI-TWPA amplifier. This runs counter to the aim of the study presented here. Another possible solution is to pattern the central line as a series of wider and narrower sections (figure \ref{fig:wider}). The narrower sections keep $I_*$ low while the wider sections decrease the number of squares per cell which, in turn, decreases the inductance per cell. Under this condition, following equation \ref{eq:kin}, the total kinetic inductance is composed of two contributions, $L_k=L_{k_{W}}+L_{k_{N}}$ (with $W$ for wide and $N$ for narrow). However, since the current $I_*$ is proportional to the line width, if $w_{W}>w_{N}$, then $I_{*W}^2 >> I_{*N}^2$ and as a result, $L_k\simeq L_{k_{N}}$ is the dominant contribution. Preliminary measurements performed with a tunable resonator~\cite{Vissers2015} showed a scaling current of around 3\,mA for $w=1\,\mu$m and around 40\,mA for $w=13\,\mu$m. These values largely satisfy the above-mentioned condition.

With this solution,  the $Z_0=50\,\Omega$ condition can be satisfied with shorter stub lengths. In order to find the best configuration we simulated three different configurations, as reported in figure \ref{fig:wider}. These configurations differ by the length of the wider parts: in the configuration $(a)$ this length is $4\,\mu$m and covers one stub, in $(b)$ the length is $8\,\mu$m and covers two stubs, and, finally, in $(c)$ the length is $13\,\mu$m and covers three stubs. The goal of the performed simulations was to vary $w$ and $\ell$ for each configuration in order to find the proper $L$ and $C$ values per cell such that $Z_0=50\,\Omega$, but with stub length $\ell$ around $100\,\mu$m. The width of the narrower sections was fixed at $1\,\mu$m while the width of the wider sections varied in $7-23\,\mu$m range.           

\begin{figure}[!t]
\centering \includegraphics[width=0.45\textwidth,clip]{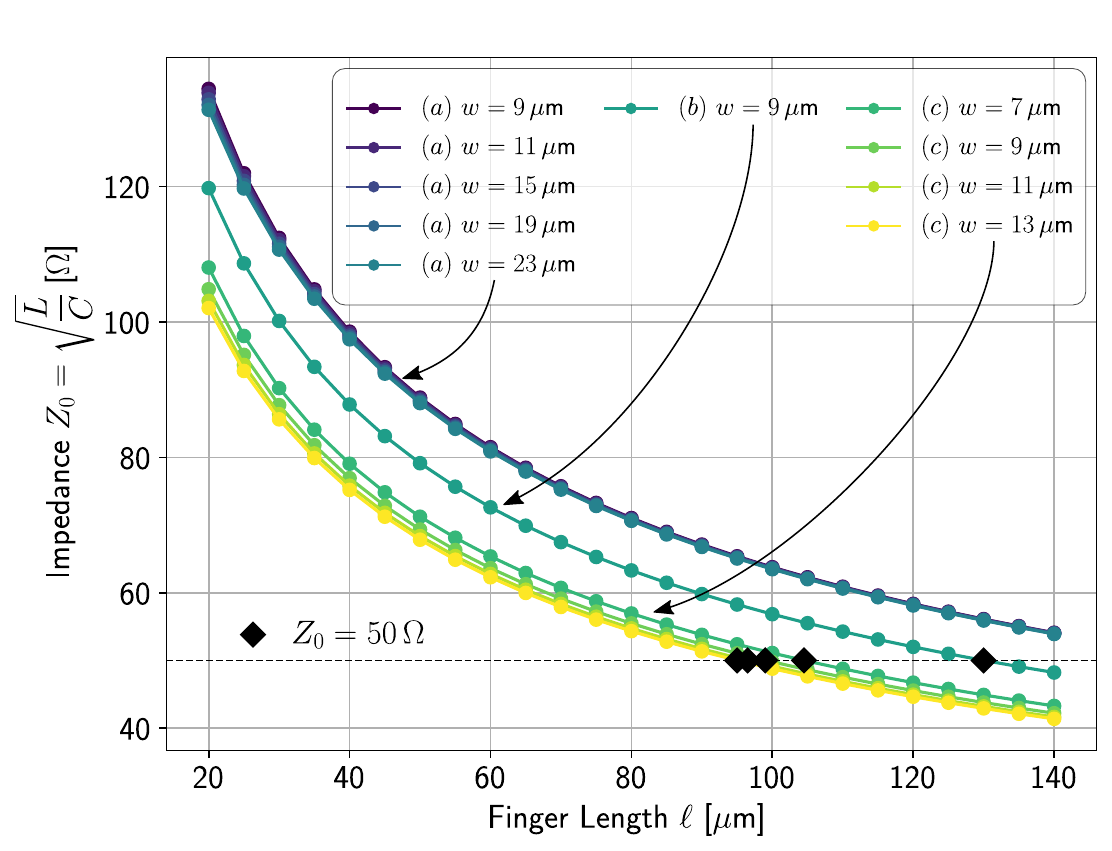}
\caption{Simulation results of line characteristic impedance $Z_0$ as a function of the stub's length $\ell$ for the configurations $(a)$, $(b)$, and $(c)$ reported in figure \ref{fig:wider}.}
\label{fig:Z0len}
\end{figure}

In figure \ref{fig:wider} the three different configurations with the current density map are shown. The characteristic impedance $Z_0$ as a function of $w$ and $\ell$ obtained from simulations is reported in figure \ref{fig:Z0len}. For the configuration $(a)$ we found that no value of $\ell$ and $w$ satisfies the $Z_0$ matching condition. In fact, despite the larger trace width $w$, simulations show that the current remains concentrated in the central part of the conductor as is shown in in figure \ref{fig:wider}$(a)$. Hence the effective width of the central conductor remains smaller than drawn and the kinetic inductance contribution of the line is too high to compensate. The resulting inductance per cell remains high and the stub length needed for matching $50\,\Omega$ falls outside our range of interest (figure \ref{fig:Z0len}). For the configurations $(b)$ and $(c)$ the current density spreads more uniformly in the wider sections (figure \ref{fig:wider}$(b)$ and \ref{fig:wider}$(c)$), the inductance per cell decreases, and it is possible to find a set of stub lengths with a capacitance $C$ per cell such that $Z_0=50\,\Omega$. Considering these simulation results we selected the configuration $(c)$ with stub length $\ell=103\,\mu$m. This length value is very close to the ones used for the previous 20\,nm version. This allowed us to design the new amplifiers using the same double spiral topology within the same chip size and with a similar number of cells. The designed devices have been recently produced at the NIST microfabrication facilities and are currently under test.  

\section{Conclusion}
In this work we produced and characterized different NbTiN films in order to investigate the thickness dependence of their intrinsic properties as the critical temperature $T_c$, the normal resistivity $R_n$, the superconducting kinetic inductance $L_k$, and the material internal loss $1/Q_i$. By exploiting DC and RF characterizations we found that the kinetic inductance increase with lowering the thickness. Starting from the results obtained for NbTiN film 10\,nm thick we have designed and simulated a new KI-TWPA layout that would provide the same gain, bandwidth, and noise performances of the previous 20\,nm version but with lower pump power, thanks to the larger kinetic inductance non-linearity. These new devices have been recently produced and are currently under test at the novel sensor laboratory at NIST.  

\section*{Acknowledgment}
This work is supported by DARTWARS, a project funded by the European Union’s H2020-MSCA Grant Agreement No. 101027746 and by the Italian Institute of Nuclear Physics (INFN) within the Technological and Interdisciplinary Research Commission (CSN5). This work is also supported by NIST through the Program on Scalable Superconducting Computing.

\bibliographystyle{IEEEtran}
\bibliography{asc2022}

\end{document}